\documentclass[conference]{IEEEtran}
\IEEEoverridecommandlockouts
\usepackage{cite}

\usepackage{amsmath,amssymb,amsfonts}
\usepackage[ruled,norelsize,vlined,linesnumbered]{algorithm2e}
\usepackage{multirow}
\usepackage{graphicx}
\usepackage{textcomp}
\usepackage{xcolor}
\usepackage{caption} 
\usepackage{url}
\usepackage{cite}
\usepackage{hyperref}
\usepackage{array}
\usepackage{textcomp}
\usepackage{stfloats}
\usepackage{verbatim}
\usepackage{subfigure}

\def\BibTeX{{\rm B\kern-.05em{\sc i\kern-.025em b}\kern-.08em
    T\kern-.1667em\lower.7ex\hbox{E}\kern-.125emX}}
\begin{document}
	
\title{ApproxPilot: A GNN-based Accelerator Approximation Framework\\
\vspace{-0.7em}
}


\makeatletter
\newcommand{\linebreakand}{%
\end{@IEEEauthorhalign}
\hfill\mbox{}\par
\mbox{}\hfill\begin{@IEEEauthorhalign}
}
\makeatother

\author{\IEEEauthorblockN{Qing Zhang}
	\IEEEauthorblockA{\textit{Institute of Computing Technology} \\
		\textit{Chinese Academy of Sciences}\\
		Beijing, China \\
		zhangqing22s@ict.ac.cn}
	\and
	\IEEEauthorblockN{Cheng Liu}
	\IEEEauthorblockA{\textit{Institute of Computing Technology} \\
		\textit{Chinese Academy of Sciences}\\
		Beijing, China \\
		liucheng@ict.ac.cn}
	\and
	\IEEEauthorblockN{Siting Liu} 
	\IEEEauthorblockA{\textit{School of Information Science and Technology} \\
		\textit{ShanghaiTech University}\\
		Shanghai, China \\
		 liust@shanghaitech.edu.cn}
	\linebreakand
	\IEEEauthorblockN{Yajuan Hui}
	\IEEEauthorblockA{\textit{Institute of Automation} \\
		\textit{China University of Geoscience}\\
		Wuhan, China \\
		 huiyj@cug.edu.cn}
	\and
	\IEEEauthorblockN{Huawei Li}
	\IEEEauthorblockA{\textit{Institute of Computing Technology} \\
		\textit{Chinese Academy of Sciences}\\
		Beijing, China \\
		lihuawei@ict.ac.cn}
	\and
	\IEEEauthorblockN{Xiaowei Li}
	\IEEEauthorblockA{\textit{Institute of Computing Technology} \\
		\textit{Chinese Academy of Sciences}\\
		Beijing, China \\
		lxw@ict.ac.cn}
}

\maketitle

\begin{abstract}
	A typical optimization of customized accelerators for error-tolerant applications such as multimedia, recognition, and classification is to replace traditional arithmetic units like multipliers and adders with the approximate ones to enhance energy efficiency while adhering to accuracy requirements. However, the plethora of arithmetic units and diverse approximate unit options result in an exceedingly large design space. Therefore, there is a pressing need for an end-to-end design framework capable of navigating this intricate design space for approximation optimization. Traditional methods relying on simulation-based or blackbox model evaluations suffer from either high computational costs or limitations in accuracy and scalability, posing significant challenges to the optimization process. In this paper, we propose a Graph Neural Network (GNN) model that leverages the physical connections of arithmetic units to capture their influence on the performance, power, area (PPA), and accuracy of the accelerator. Particularly, we notice that critical path plays a key role in node feature of the GNN model and having it embedded in the feature vector greatly enhances the prediction quality of the models. On top of the models that allow rapid and efficient PPA and accuracy prediction of various approximate accelerator configurations, we can further explore the large design space effectively and build an end-to-end accelerator approximation framework named ApproxPilot to optimize the accelerator approximation. Our experimental results demonstrate that ApproxPilot outperforms state-of-the-art approximation optimization frameworks in both performance and hardware overhead with the same accuracy constraints. 
\end{abstract}

\begin{IEEEkeywords}
	Approximate Accelerator, GNN, Approximation Optimization Framework, Design Space Exploration
\end{IEEEkeywords}

\IEEEpeerreviewmaketitle

\section{Introduction}

Approximate computing has emerged as a powerful computing paradigm that leverages application-tolerable inaccuracies to achieve greater energy efficiency and performance with minimal impact on output quality\cite{jiang2020approximate,xu2015approximate,mittal2016survey,mahdiani2009bio,shin2011new}. It can be applied in a broad domains of applications such as machine learning, computer vision, and signal processing. A prevalent method to take advantage of approximate computing is to substitute precise arithmetic units in accelerators with their approximate without altering the fundamental accelerator architecture. However, application-specific accelerators often contain dozens to hundreds of arithmetic units, each unit providing a number of approximation options that present unique balances between accuracy, performance, and overhead. Consequently, the approximation design space is exceedingly large, rendering manual determination of optimal configurations for all arithmetic units nearly infeasible. Therefore, an end-to-end framework for efficient approximation design space exploration becomes imperative.

To tackle the challenge of approximation optimization of accelerators, several methods \cite{nepal2014abacus, mrazek2019autoax} have been proposed to search through an approximate arithmetic unit library, aiming to pinpoint the best approximation option for each unit in the approximate accelerator. ABACUS\cite{nepal2014abacus}, a pioneering initiative, employs an evolutionary strategy based on the non-dominated sorting genetic algorithm (NSGA-II) to substitute approximate operators within the abstract syntax tree of a register transfer level (RTL) design. However, it evaluates the quality metrics such as area and latency by extracting performance, power, and area (PPA) from time-consuming CAD tools, which limits the design space exploration (DSE) efficiency substantially and affects the quality of results eventually. In contrast, AutoAX \cite{mrazek2019autoax} leverages machine learning techniques such as random forests to estimate the PPA and accuracy of various approximate accelerators. This approach by replacing CAD simulations with trained models greatly reduces the evaluation overhead and enables more intensive DSE accordingly. Despite the fast evaluations with traditional machine learning models \cite{mrazek2019autoax,prabakaran2023xel,dai2018fast}, they typically regard the accelerator as a black-box function of the approximate arithmetic units. These evaluation models may suffer limited evaluation accuracy, especially as the architectural complexity of the accelerator increases or when dealing with a large number of arithmetic units.

Inspired by the success of graph neural network (GNN) based PPA modeling of large-scale circuits \cite{wu2021ironman,wu2022high,li2022noception,wu2022ironman,10416120}, the connection topology of the arithmetic units is considered in the evaluation models and GNNs are employed to enhance the PPA and accuracy prediction for accelerators configured with various approximate arithmetic units. The baseline GNN models typically can deliver accurate predictions of power and area, alongside satisfactory accuracy predictions, yet they fall short in latency estimations when compared to simulated results. This discrepancy arises because latency is governed by the accelerator's critical path, which is part of the entire accelerator graph topology, and with basic graph information, it is not possible to explicitly distinguish between the critical path and other paths in the graph.
 
To address this, we incorporate the critical path as a novel feature dimension in our GNN models. As the critical path is unknown during the prediction, we propose an independent GNN model to predict whether an arithmetic is part of the critical path. Eventually, we propose a two-step GNN model for the latency prediction. Further, we apply several pruning strategies to condense the design space and pre-processing the dataset for model training. Finally, we propose ApproxPilot, an end-to-end optimization framework for generic approximation of hardware accelerators. According to our experiments on three computation-intensive accelerators, ApproxPilot remarkably surpasses existing frameworks, producing superior Pareto-frontier curves under diverse conditions.

The major contributions of this paper are as follows: 
\begin{itemize}
\item We develop a GNN model to predict the PPA and accuracy of approximate accelerators with various approximate arithmetic units for the first time.  Particularly, we observe the critical path information plays a key role in the latency prediction and propose a two-step GNN model to incorporate this information for the PPA and accuracy prediction, significantly improving the modeling quality and scalability.

\item On top of the proposed PPA and accuracy models, we develop an end-to-end framework ApproxPilot, covering dataset construction, PPA and accuracy modeling, and DSE, for approximation optimization of generic hardware accelerators. This framework is open sourced on GitHub \footnote{\url{https://anonymous.4open.science/r/840c8c57-3c32-451e-bf12-0e243562/}}.

\item Our experimental evaluation across a suite of accelerator benchmarks, including the Sobel edge detector, Gaussian filter, and KMeans operator, shows that ApproxPilot achieves significant reduction in chip area and power consumption compared to the state-of-the-art frameworks while maintaining the same accuracy constraints.

\end{itemize}

\section{Related Work}
\subsection{Approximate Circuit Design and Optimization}
 
Approximate computing leverages the inherent fault tolerance in various applications, such as signal processing, image processing, and data mining, to enhance computing efficiency by reducing substantial hardware resources while incurring only a minor accuracy loss. A wealth of research has been conducted to investigate approximate computing from multiple perspectives, as detailed in recent surveys \cite{scarabottolo2020approximate,jiang2020approximate,xu2015approximate,mittal2016survey,scarabottolo2018circuit}. This section will primarily discuss circuit-level approximation techniques, focusing on the approximation of arithmetic units and the use of these arithmetic units for accelerator approximation.

Circuit approximation typically involves substituting accurate operations with their approximate counterparts. There has been extensive literature on designing approximate arithmetic units, including multipliers and adders \cite{brent1982regular,han1987fast,kahng2012accuracy,liu2015design,rehman2016architectural,saadat2019approximate,shafique2015low,kulkarni2011trading}, which are beneficial to accelerator performance. Automatic design tools have also been examined to streamline the creation of various approximate arithmetic units, offering different trade-offs between the quality of result (QoR) and energy efficiency. Moreover, libraries of approximate arithmetic units have been established, for example Evoapprox8b\cite{mrazek2017evoapprox8b}, SMApproxlib\cite{ullah2018smapproxlib}, featuring a range of units with diverse approximation settings to cater to diverse accelerator and application requirements.

In addition to design approximate arithmetic units, selecting the optimum approximation for each unit within a customized accelerator is paramount but presents a considerable challenge due to the immense design space. To date, this issue has not been thoroughly addressed in existing literature. ABACUS\cite{nepal2014abacus} is noted as the foundational study in this area, utilizing the time-consuming CAD tools to assess the PPA of accelerators, thus significantly constraining DSE efficiency. Particularly, fast and precise evaluation of PPA and accuracy of accelerators with distinct approximate arithmetic units is imperative for efficient DSE and remains a significant impediment to approximate accelerator optimization. To alleviate the burden of costly PPA and accuracy assessments, AutoAX\cite{mrazek2019autoax} employs a random forest method to address the time-consuming CAD analysis. In spite of the rapid evaluation, it is limited to designs with a small number of arithmetic units and becomes less accurate for larger designs. Therefore, there is an acute need for more efficient evaluation algorithms.

\subsection{PPA and Accuracy Prediction Algorithms}
Evaluating the PPA and accuracy with CAD tools simulations can be extremely slow and expensive. Hence, PPA and accuracy prediction based on model construction becomes crucial for circuit and architecture optimization and has been widely researched. While black-box modeling represented by machine learning offers prediction speed, but it lacks the accuracy needed for effective DSE in general. Therefore, a white-box or grey-box modeling approach that incorporates more detailed circuit and architectural information could be beneficial. GNN models that are capable to map the structural details of circuits to the connectivity relations between nodes provide greater precision than traditional machine learning methods and have been successfully applied to many hardware  architecture modeling problems\cite{wu2022ironman, 10131995, 10411398, 9401612, 10476462,10416120}. Most of them \cite{ustun2020accurate, wu2022high, li2022noception, wu2022ironman, wu2021ironman} are aimed at netlists, architectural parameters \cite{li2022noception}, and High-Level Synthesis (HLS) pragmas \cite{wu2022ironman}. But, predicting PPA and accuracy for accelerators configured with different approximate arithmetic units remains a distinct and challenging issue that is often neglected.

\section{ApproxPilot Framework}
In this study, we propose ApproxPilot, an end-to-end framework for approximation optimization of customized accelerators, as presented in Figure \ref{fig:ApproxPilot-Overview}. Given the accelerator design and user constraints, ApproxPilot autonomously selects the most suitable approximate arithmetic units from an established approximate library for each arithmetic unit in the customized accelerator. This selection facilitates both enhanced performance and energy efficiency while adhering to the user constraints, such as accuracy and hardware overhead constraints. As the design space exploration necessitates numerous, time-consuming quality evaluations, such as PPA and accuracy for accelerators configured with various approximate units. ApproxPilot employs GNN models rather than CAD tools to obtain the evaluation results. These models consider the topology of interconnected arithmetic units in accelerator, enabling both fast and precise evaluations of accelerator quality. With these efficient evaluation models, ApproxPilot conducts thorough design space exploration and generates superior approximate accelerators. The core procedures of ApproxPilot, including design space pruning, accelerator quality evaluation, and design space exploration, are detailed further in the rest of this section.

\begin{figure}[tb] 
    \centering
    \includegraphics[width=1\linewidth]{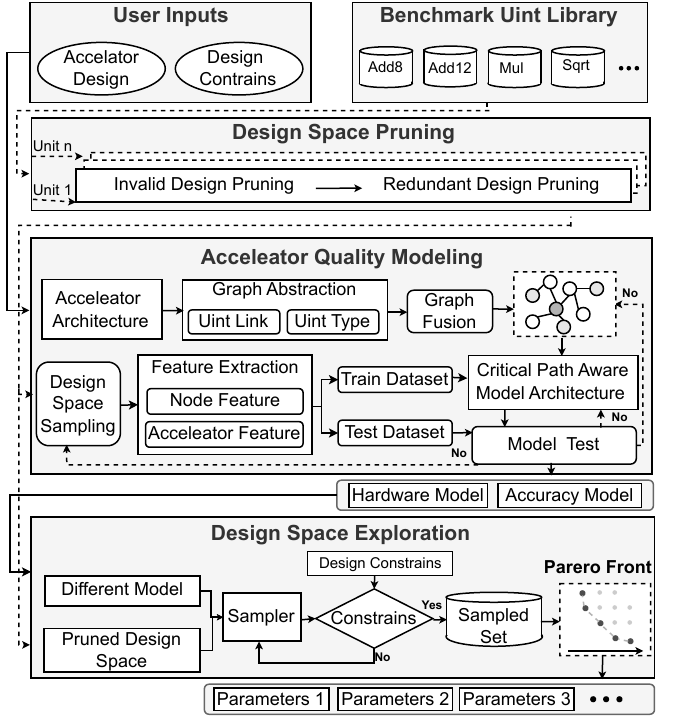} 
    \caption{ApproxPilot Overview}
    \vspace{-1.5em}
    \label{fig:ApproxPilot-Overview}
\end{figure}

\subsection{Design Space Pruning}
ApproxPilot relies on an extensive library of approximate computing unit candidates to fulfill the needs of the various approximate accelerator designs. However, existing approximate arithmetic libraries are still inadequate and cannot fully cover the requirements of the approximate accelerators. which mainly consists of multiplier and adders operations, but in some accelerators, sqrt and sub are indispensable. In this work, we build a more comprehensive library by combining different libraries. In addition, the number of approximate candidates of different computing units can vary dramatically. Some of them may include hundreds or even thousands of candidates, which will induce an extremely large design space and complicate the optimization procedure. Particularly, the candidates can be very close to each other and some of them may contribute little to the approximation optimization. 

To address this problem, we opt to prune the design space and alleviate the optimization efforts. The basic idea is to build a feature vector for each arithmetic unit candidate and squeeze the invalid or redundant candidates based on the feature vectors. For each arithmetic unit candidate, it can be characterized with a vector $V = [MSE, Area, Power, Latency]$, where the MSE metric stands for the errors between approximate outputs and accurate outputs on all possible inputs, and the PPA is extracted from synthesis report. In general, we have two different pruning strategies applied to the approximate arithmetic library.

\subsubsection{Invalid Design Pruning} Some of the arithmetic units in the library exhibit lower precision and performance but increased area or power requirements as formulated in Equation \ref{eq:Circuit-Design-Redundancy}. Specifically, $l$ represents the different dimensions of the $V$ vector.Basically, the arithmetic unit characterized with $V_{i}$ can be completely replaced by $V_{j}$ and thus, $V_{j}$ can be removed from library.
\begin{equation} \label{eq:Circuit-Design-Redundancy}
   {\forall} V_{i} ,V_{j}  \in N,V_{i}^{l} \geq V_{j}^{l} \text { for } l=1, \ldots, 4
\end{equation}


\subsubsection{Redundant Design Pruning}To avoid redundancy in the design space caused by approximate units with similar performance,We employ the K-means algorithm to evaluate the distance between feature vectors of two arithmetic units $V_{i}$ and $V_{j}$ in the candidate unit library.The distance is calculated with Equation \ref{eq:vec-dist}, where $\theta$ represents the distance threshold and $\rho$ denotes the normalization coefficients of each feature in the V vector. If the distance $\theta$  between two candidate units is lower than this threshold, one of them will be randomly selected for pruning.
\begin{equation} \label{eq:vec-dist}
  {\forall} V_{i} ,V_{j}  \in N, \sqrt{\sum_{i=3}^{i=0}(\rho_{i} (V_{i}^{l}-V_{j}^{l}))^{2}}\le \theta  
\end{equation}
\begin{figure*}[h] 
    \centering
    \includegraphics[width=0.8\linewidth]{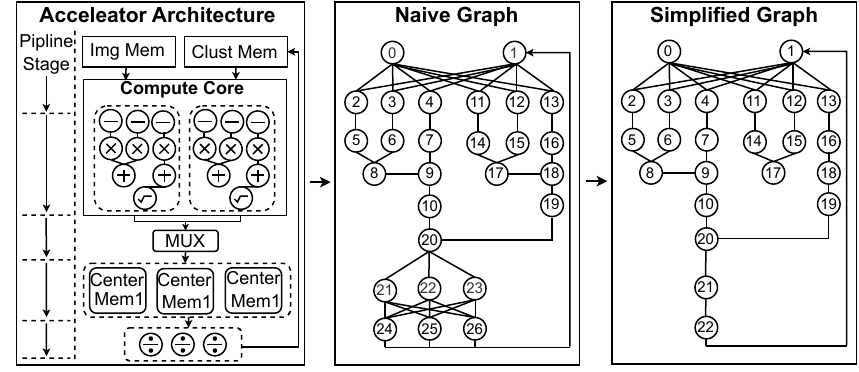} 
    \caption{Graph Abstraction}
    \label{fig:graph-abstraction}
\end{figure*}

\subsection{PPA and Accuracy Modeling}
To search for the optimized approximation configuration over the vast design space, ApproxPilot requires a large number of PPA and accuracy evaluation of accelerators with different approximation configurations. The evaluation needs to be fast and precise to ensure extensive and effective design space exploration. Unlike simulation-based methods, which are often slow and costly, we adopt model-based prediction due to its rapidity and have it detailed in the following subsections.

\subsubsection{Dataset Construction} \label{sec:dataset}
In order to obtain efficient prediction models, we start by building a dataset that represents the design space of the approximate accelerators through random sampling. The dataset includes diverse configurations of approximate arithmetic units. Notably, to enhance the dataset's representativeness and avoid redundancy, we examine the accelerators with symmetric structures and eliminate duplicate samplings of equivalent designs. When the sampling is determined, we further evaluate the quality features of each approximate accelerator sample. The quality features mainly consist of two aspects: unit features and accelerator features. For unit features, the focus is on efficiently constructing feature vectors for each node in the graph. Accelerator features primarily entail architectural information about approximation accelerators, including latency, area, power, and accuracy. While the former three can be drawn from the synthesis report of commercial EDA tools such as Synopsys Design Compiler, the accuracy is characterized with errors relative to precise outputs using Structural Similarity Index Measure (SSIM) for image output and Mean Squared Error (MSE) as recommended by AxBench~\cite{yazdanbakhsh2016axbench}. Furthermore, to bypass the slow RTL simulation for accuracy evaluation, we opt for high-level functional accelerator models for faster output generation. Ultimately, the combination of these features forms a dataset configuration for each accelerator sample, which serves to train and validate the prediction models. Following machine learning practices, we partition the dataset into 90\% for training set and 10\% for testing set.
\subsubsection{Prediction Model Construction}
Unlike prior approaches that view the accelerator as a black box, we consider not only the approximate arithmetic units feature but also their physical connections. We employ a GNN model to capture the impacts of these units on the accelerator’s PPA and accuracy. Although applying baseline GNN model and basic unit feature to predict PPA and accuracy presents challenges due to complex node feature characterization and graph construction, our method addresses these effectively.

Graph construction in our model treats each approximate arithmetic unit as a node, with physical connections as edges. Fixed components of the accelerator, unaffected by the approximation choices, are abstractly modeled based on their function and merged when they share the same input-output relationships to maintain connection integrity while reducing graph complexity. Node feature characterization includes all critical factors affecting the accelerator's PPA and accuracy, alongside multiple error metrics to thoroughly depict the accuracy of arithmetic units. Additionally, we differentiate node types and approximation levels, encoding enumeration data into the feature vector for completeness. Figure \ref{fig:graph-abstraction} illustrates an example of graph construction based on Kmeans filter. Arithmetic units that can be optimized with approximate units are modeled as independent nodes. Center Mem1, Center Mem2, and Center Mem3 are memory components and they are modeled with nodes in the first step. While the three memory blocks are fixed components and they have the same incoming nodes and outgoing nodes, they are merged to be a single one in the second stage.  Similarly, division operators are fixed and merged as well. In contrast, img-mem and cluster-mem that have different inputs remain independent nodes. 

Node feature characterization necessitates the inclusion of all major factors influencing the PPA and accuracy of approximate accelerators as listed in Table \ref{tab:features}. Thus, PPA and accuracy of the approximate arithmetic units that pose direct influence on the overall accelerator is utilized as node features. While PPA independent with inputs is relatively stable, accuracy depends on realistic inputs is more difficult to measure. In this work, we employ combined error metrics including Mean Absolute Error (MAE), Mean Relative Error (MRE), Mean Square Error (MSE), and Worst Case Error (WCE) instead. Moreover, we have node type to categorize the components within the accelerator and utilize the approximation level to indicate the degree of each unit’s approximation. Both node type and approximation level are encoded into the feature vector as binary data. 

\begin{table}[h]
\centering
\Large
\caption{Node features and example values.}
\label{tab:features}
\resizebox{\linewidth}{!}{
\begin{tabular}{c|c|c|c}
\hline
&  Node Feature  &  Description  &  Value       \\ \hline
\multirow{4}{*}{\begin{tabular}[c]{@{}c@{}}PPA \\ Type\end{tabular}} 
&  Area  &  Chip area of the approximate arithmetic unit &  Float Value \\
&  Power &  Power of the approximate arithmetic unit &  Float Value \\
&  Latency &  Latency of the approximate arithmetic unit &  Float Value \\
&  Compute Type &  \begin{tabular}[c]{@{}c@{}}the difference between control unit/memory\\ /compute unit(add, sub, mul, sqrt)\end{tabular} &  One-Hot  \\ \hline
 Status &  On critical path? &  Is the node located on the critical path? &  Boolean \\ \hline
\multirow{4}{*}{\begin{tabular}[c]{@{}c@{}}Accuracy \\ 
(SSIM)\end{tabular}}
&  MAE &  Mean Absolute Error &  Float Value \\
&  MRE &  Mean Relative Error &  Float Value \\
&  MSE &  Mean Square Error &  Float Value \\
&  WCE &  Worst-Case Relative Error &  Float Value \\ \hline
\end{tabular}
}
\end{table}

With both graph structure and node features, we can train a Graph Neural Network (GNN) model to predict the PPA and accuracy of accelerators equipped with different approximate arithmetic units. It is crucial to note that latency, the key performance metric, relies on the accelerator's critical paths rather than the timing paths of the entire design, which differs from power, area, and accuracy metrics. When PPA and accuracy predictions are integrated into a single GNN model, it fails to capture the influence of critical paths, and latency predictions diverge noticeably from simulated outcomes. To address this problem, we introduce a second GNN model focused exclusively on predicting critical paths at the granularity of the arithmetic units. Given that critical path information is also obtainable from synthesis reports, this addition does not necessitate further data construction effort. With the critical path predictions, nodes are categorized based on their location within these paths, allowing critical path data to be incorporated into the feature vector and, consequently, improve latency prediction precision. As demonstrated in Figure \ref{fig:GNN_Model}, the prediction model operates in a two-stage process:(1) Node-level Classification: the first stage predicts critical path and integrates this information into node features, (2) Graph-Level Regression: the second stage uses these enhanced features for PPA and accuracy prediction. 
\begin{figure}[h] 
    \centering
    \includegraphics[width=1\linewidth]{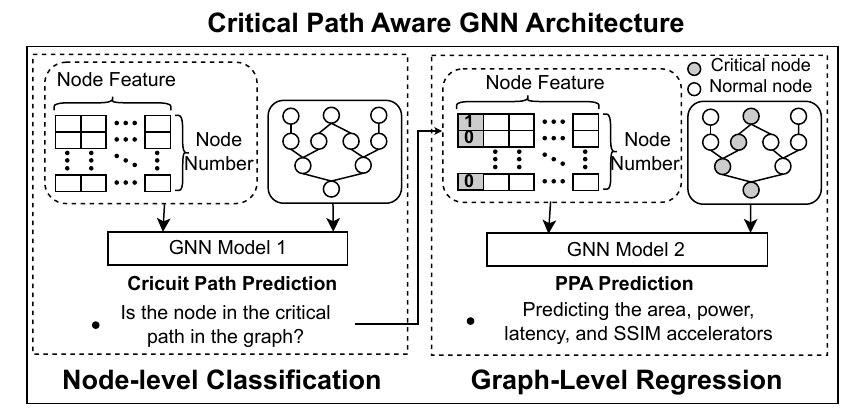} 
    \caption{Critical Path Aware Model Architecture}
    \label{fig:GNN_Model}
\end{figure}

\vspace{-1em}
\subsection{Design Space Exploration}
With the GNN model, we can obtain PPA and accuracy of an approximate accelerator rapidly without using the time-consuming simulation. In this case, we can further construct the Pareto curve to guide the approximate accelerator optimization under different user constraints. Our optimization goal is to find the smallest resource-cost accelerator design configuration that satisfies user constraints from the vast design space.  The key of this step is the selection of the sampling function. There have been many different algorithms proposed to build Pareto curves over a large design space and we evaluate a few representative methods in this work. NSGAIII\cite{deb2013evolutionary} sampling that outperforms the others are utilized in the Design Space Exploration.

For the design space exploration stage,initially, we randomly select some accelerator designs from the pruned design space as the initial parent population $P_{0}$. Then, by performing operations such as "crossover", "mutation", and "recombination" on some units within the parent population $P_{0}$, we create a descendant population $Q_{0}$. Next, using the trained GNN model, we evaluate $P_{0}$ and $Q_{0}$ to obtain their PPA and accuracy metrics. Subsequently, non-dominated sorting is applied to the parent population $P_{0}$ and descendant population $Q_{0}$, and the crowding distance of each sample is calculated. Afterwards, appropriate accelerator designs are selected to form a new parent population $P_{i}$ based on user design constraints for iteration. After $N$ iterations, we will output the parent population set $P$ for constructing the Pareto front. In order to prevent the optimization process of DSE from being trapped in local optima, when it is observed that the parent population $P_{i}$ remains unchanged, we will restart by conducting random sampling from the design space for future iterations.

\section{Experiment Results}
In this section, we evaluate ApproxPilot for approximate accelerator design and compare with AutoAX, a state-of-the-art design framework \cite{mrazek2019autoax}. Additionally, we investigate the key techniques employed in ApproxPilot and substantiate their benefits through extensive experimental validation.

\subsection{Experiment Setups}
\textbf{Benchmark:} In this experiment, we utilize Sobel edge detector, Gaussian filter, and kmeans clustering as the basic approximate accelerators. The Kmeans clustering algorithms are selected from AXBench \cite{yazdanbakhsh2016axbench}, whereas the Sobel edge detector and Gaussian filter is taken from the benchmarks used in AutoAX \cite{mrazek2019autoax}. The basic operation configurations for these accelerators are summarized in Table \ref{tab:opt-summary}. For accuracy assessment, we utilize the BSD500\cite{arbelaez2010contour} dataset as inputs and quantify the differences between outputs from precise and approximate accelerators. Specifically, the average Structural Similarity Index Measure (SSIM) is utilized as the accuracy metric and it is obtained with a Python-based accelerator model. The PPA metrics are extracted from the synthesis report generated by Synopsys Design Compiler using a 45-nm technology.

\begin{table}[h]
\caption{Operator summary of Sobel edge detector, Gaussian filter, and Kmeans Clustering.}
\Large
\resizebox{\linewidth}{!}{
\begin{tabular}{ccccccccl}
\hline
\multirow{2}{*}{\textbf{Application}} & \multicolumn{3}{c}{\textbf{Adder}}           & \textbf{Subtractor}         & \multicolumn{2}{c}{\textbf{Multiplier}} & \textbf{Sqrt}               & \textbf{Total}         \\
                                      & 8-bit & 12-bit & \multicolumn{1}{c|}{16-bit} & \multicolumn{1}{c|}{10-bit} & 8-bit   & \multicolumn{1}{c|}{8*4bit}   & \multicolumn{1}{c|}{18-bit} & \multicolumn{1}{c}{\#} \\ \hline
\multicolumn{1}{c|}{Sobel}            & 2     & 2      & \multicolumn{1}{c|}{-}      & \multicolumn{1}{c|}{1}      & -       & \multicolumn{1}{c|}{-}        & \multicolumn{1}{c|}{-}      & 5                      \\
\multicolumn{1}{c|}{Gaussian}         & -     & -      & \multicolumn{1}{c|}{8}      & \multicolumn{1}{c|}{-}      & -       & \multicolumn{1}{c|}{9}        & \multicolumn{1}{c|}{-}      & 17                     \\
\multicolumn{1}{c|}{Kmeans}           & -     & -      & \multicolumn{1}{c|}{2}      & \multicolumn{1}{c|}{6}      & 6       & \multicolumn{1}{c|}{-}        & \multicolumn{1}{c|}{2}      & 16                     \\ \hline
\end{tabular}
}
\label{tab:opt-summary}
\end{table}


\textbf{Approximate Operator Library:}
The approximate operator library provides candidate approximate operators required by the accelerators and is the basis of ApproxPilot. In this work, we selected EvoApproxLib \cite{mrazek2017evoapprox8b} as the basic approximate unit library and augmented the library to fulfill all the arithmetic unit requirements of the accelerators. Specifically, we apply the same methodology used for EvoApproxLib to generate a broader array of approximate operators, such as Add8u and Add16u. In addition, for the approximate operators absent in the library, we employed CGP\cite{vasicek2014evolutionary}, which was also used for EvoApproxLib's generation \cite{mrazek2017evoapprox8b}, to create additional operators, including sub and sqrt operations. The approximate operators that are available in our initial library are summarized in Table \ref{tab:opt_num_list}.

\begin{table}[h]
\caption{Approximate arithmetic operator library}
\Large
\resizebox{\linewidth}{!}{
\begin{tabular}{cccccccc}
\hline
\multirow{2}{*}{\textbf{Instance}} & \multicolumn{3}{c}{\textbf{Adder}}           & \textbf{Subtractor}         & \multicolumn{2}{c}{\textbf{Multiplier}} & \textbf{Sqrt} \\
                                   & 8-bit & 12-bit & \multicolumn{1}{c|}{16-bit} & \multicolumn{1}{c|}{10-bit} & 8-bit   & \multicolumn{1}{c|}{8*4bit}   & 18-bit        \\ \hline
\multicolumn{1}{c|}{number}        & 31    & 26     & \multicolumn{1}{c|}{21}     & \multicolumn{1}{c|}{12}     & 35      & \multicolumn{1}{c|}{32}       & 7             \\ \hline
\end{tabular}
}
\label{tab:opt_num_list}
\end{table}

\textbf{Design Framework Setups:} 
In order to train the PPA and accuracy prediction models of approximate accelerators, we selected 55,000, 105,000, and 105,000 samples randomly from the pruned design space of Sobel Edge Detector, Gaussian Filter, and Kmeans clustering, respectively. We partitioned each dataset into distinct subsets, allocating 90\% for training, 10\% for testing. To be consistent with the methodologies proposed in AutoAX\cite{mrazek2019autoax}, we adopted random forest model for the PPA and accuracy prediction. For ApproxPilot, we examine a set of GNN models and select the most effective GNN model for the prediction. Each GNN model is configured with five layers and a hidden dimension of 300. For the GNN training, we set the batch size to be 5 and the initial learning rate be 0.001. The training was conducted over 100 epochs with the Adam optimizer. The learning rate, dropout rate, and the number of layers were dynamically tuned according to the test dataset performance.

\subsection{Overall Approximation Optimization}
We apply ApproxPilot and AutoAX to explore the approximation design spaces for the three accelerators specified in our benchmark. Subject to accuracy constraints, it is possible to target either chip area or latency for optimization. The Pareto-optimal curves, including SSIM-Area and SSIM-Latency, are depicted in Figure \ref{fig:Pareto-Curves}. The analysis reveals that ApproxPilot generally surpasses AutoAX in optimizing both area and latency across all three accelerators. The superiority is less pronounced in the Sobel Edge Detector due to its relatively confined design space and the comparable PPA and accuracy prediction quality between AutoAX and ApproxPilot. However, the advantages become more evident with the Gaussian Filter and K-means, which have larger design spaces for approximation optimization and the connection relationships with more complex computing units, as detailed in Table \ref{tab:design-space}. Notably, AutoAX's SSIM-Latency Pareto-optimal curve for the Gaussian Filter optimization is constrained to a limited range. It is because that when the design space further expands, the exploration of hill-climbing algorithms is limited to neighbor design points, which can lead to the DSE process stuck in local optima. Moreover, ApproxPilot exhibits more significant improvements in latency optimization than in area optimization, mainly attributed to the enhanced latency prediction quality of the proposed GNN models, which will be further discussed in the subsequent evaluations.

The density of sampling points along the Pareto-frontier curves is an essential metric that can distinguish between the efficacy of the different design space exploration methodologies. As indicated by the data in Table \ref{tab:op-number-on-Pareto-frontier}, ApproxPilot's Pareto frontiers feature significantly more sampling points than those generated by AutoAX, pointing to a more comprehensive exploration of the design space.

\begin{table}[h]
\centering
\Large
\caption{The number of sampling points on the Pareto frontiers constructed with ApproxPilot and AutoAX.}
\resizebox{\linewidth}{!}{
\begin{tabular}{c|c|ccc}
\hline
Method                       & Operator & Area-SSIM & Power-SSIM & Latency-SSIM \\ \hline
\multirow{3}{*}{Autoax}      & Sobel    & 24        & 32         & 21           \\
                             & Gaussian & 15        & 22         & 14           \\
                             & Kmeans   & 65        & 39         & 17           \\ \hline
\multirow{3}{*}{ApproxPilot} & Sobel    & 46      & 53        & 34          \\
                             & Gaussian & 27       & 47        & 41           \\
                             & Kmeans   & 152       & 112        & 74          \\ \hline
\end{tabular}
}
\label{tab:op-number-on-Pareto-frontier}
\end{table}

\begin{figure}[!t]
\centerline{\includegraphics[width=0.5\textwidth]{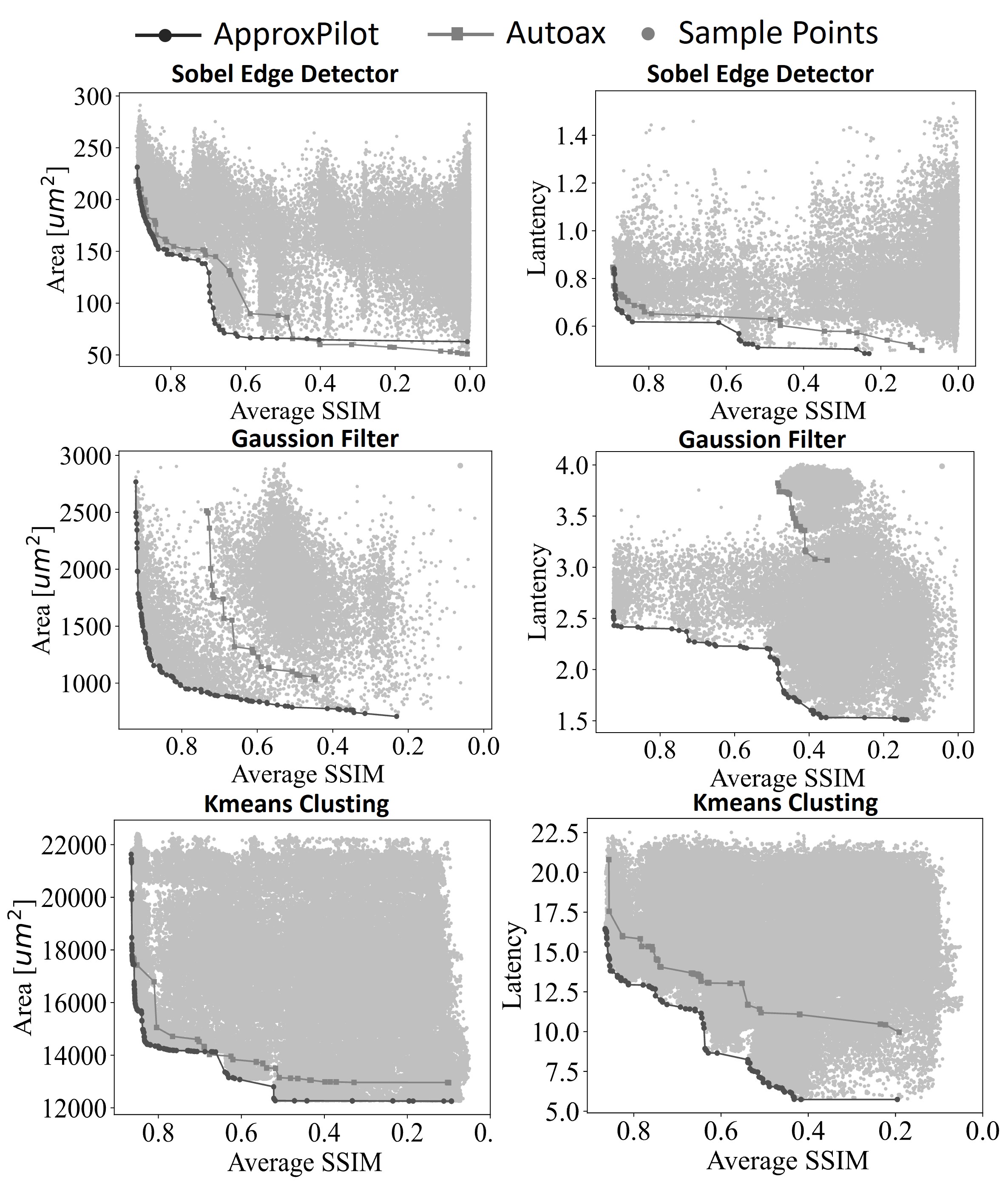}}
\caption{Pareto frontier curves generated by ApproxPilot and AutoAX.}
\label{fig:Pareto-Curves}
\end{figure}

\begin{table*}[ht]
\centering

\caption{Comparison of Autoax and ApproxPilot methods on different operators.}
\label{tab:model-comparison}
\resizebox{0.8\linewidth}{!}{
 \begin{tabular}{c|c|cccc|cccc}
\hline
\multirow{2}{*}{Method}      & \multirow{2}{*}{Operator} & \multicolumn{4}{c|}{R2\_Score}  & \multicolumn{4}{c}{MAPE}         \\ \cline{3-10} 
                             &                           & Area  & Power & Latency & SSIM  & Area   & Power & Latency & SSIM  \\ \hline
\multirow{3}{*}{Autoax}      & Sobel                     & 0.99  & 0.98  & 0.78    & 0.98  & 0.016  & 0.019 & 0.05    & 0.468 \\
                             & Gaussian                  & 0.91  & 0.86  & 0.68    & 0.72  & 0.03   & 0.05  & 0.05    & 0.22  \\
                             & Kmeans                    & 0.99  & 0.913 & 0.91    & 0.835 & 0.005  & 0.018 & 0.02    & 0.082 \\ \hline
\multirow{3}{*}{ApproxPilot} & Sobel                     & 0.99  & 0.975 & 0.98    & 0.98  & 0.0151 & 0.012 & 0.012   & 0.606 \\
                             & Gaussian                  & 0.975 & 0.94  & 0.925   & 0.956 & 0.024  & 0.03  & 0.023   & 0.142 \\
                             & Kmeans                    & 0.99  & 0.98  & 0.935   & 0.86  & 0.0034 & 0.01  & 0.019   & 0.074 \\ \hline
\end{tabular}
}
\end{table*}
\subsection{PPA and Accuracy Prediction}
\textbf{Overall prediction model comparison:} PPA and accuracy prediction is key to the overall approximation optimization design frameworks. In this experiment, we evaluate various predictive models, including the random forest model employed by AutoAX and the graph neural networks (GNNs) utilized in ApproxPilot. Table \ref{tab:model-comparison} employs R2 Score\cite{ash1999r2} and Mean Absolute Percentage Error (MAPE)\cite{tayman1999validity}, as defined in Equation \ref{eq:r2} \ref{eq:mape}.
\begin{equation} \label{eq:r2}
   R^{2}= 1-\frac{\sum_{n}^{i=1}(y_{i}-\hat{y_{i}})^{2}}{\sum_{n}^{i=1}(y_{i}-\bar{y_{i}})^{2} }  
\end{equation}
\begin{equation} \label{eq:mape}
   \text { MAPE }=\frac{100 \%}{n} \sum_{i=1}^{n}\left|\frac{\hat{y}_{i}-y_{i}}{y_{i}}\right|  
\end{equation}

To measure the quality of these models by assessing the consistency of the predictions against simulation-based results. A higher R2 Score suggests more consistent predictions, while a lower MAPE indicates better prediction quality. ApproxPilot is shown to provide superior predictions for Gaussian and K-means clustering accelerators with an increased number of arithmetic operators, aligning well with the above design space exploration. Furthermore, area and power exhibit less variability within accelerators, leading to more straightforward predictions, with both ApproxPilot and AutoAX achieving high accuracy. However, latency and SSIM are notably affected by the interconnection of arithmetic units. Hence, the random forest model in AutoAX shows limited effectiveness in comparison to ApproxPilot’s GNNs, which is aware of the connection topologies.

Meanwhile, we also tested the prediction accuracy of critical paths for three different operators. For Sobel, Gaussian, and K-means operators,the prediction accuracies of their critical paths are 91\%, 88\%, and 87\% respectively. High prediction accuracies were achieved for all three operators, because for the task of critical path prediction, GNNs can aggregate and compare features of neighboring nodes, thus distinguishing Latencys on different paths and obtaining critical path information

\textbf{Critical path awareness:} To further demonstrate the effectiveness of the proposed critical path aware two-stage GNN model, we have it compared to both a baseline single-stage GNN model and the random forest model for latency prediction on the test data of Gaussian Filter accelerator visually. As shown in Figure \ref{fig:model-comparison}, each data point stands for a accelerator design in the test data. The x-dimension and y-dimension of each data point refer to the predicted latency and simulated latency respectively. The red line represents perfect prediction. Hence, the data points that distribute more closely around the red line indicate better prediction quality in general. It is clear that the proposed critical path aware model has more data points located around the ideal line and outperforms the other two models accordingly. In contrast, the baseline GNN model is insufficient for the latency prediction and shows much larger latency variation, which confirms that the critical path information plays an important role in latency prediction.
However, both the baseline GNN model and the random forest model do not explicitly incorporate Critical path information in node encoding, resulting in low accuracy in latency prediction. On the other hand, the Critical path-aware GNN model first predicts and encodes Critical  path information using one GNN model and then concatenates it with a second GNN model to predict overall latency, thus significantly improving latency prediction accuracy. From experimental data, the Critical path-aware GNN model has increased the R2 coefficient by 25\% compared to the random forest model and by 20\% compared to the baseline GNN model.

\begin{figure*}[h]
  \centering
  \subfigure{\includegraphics[width=0.9\textwidth]{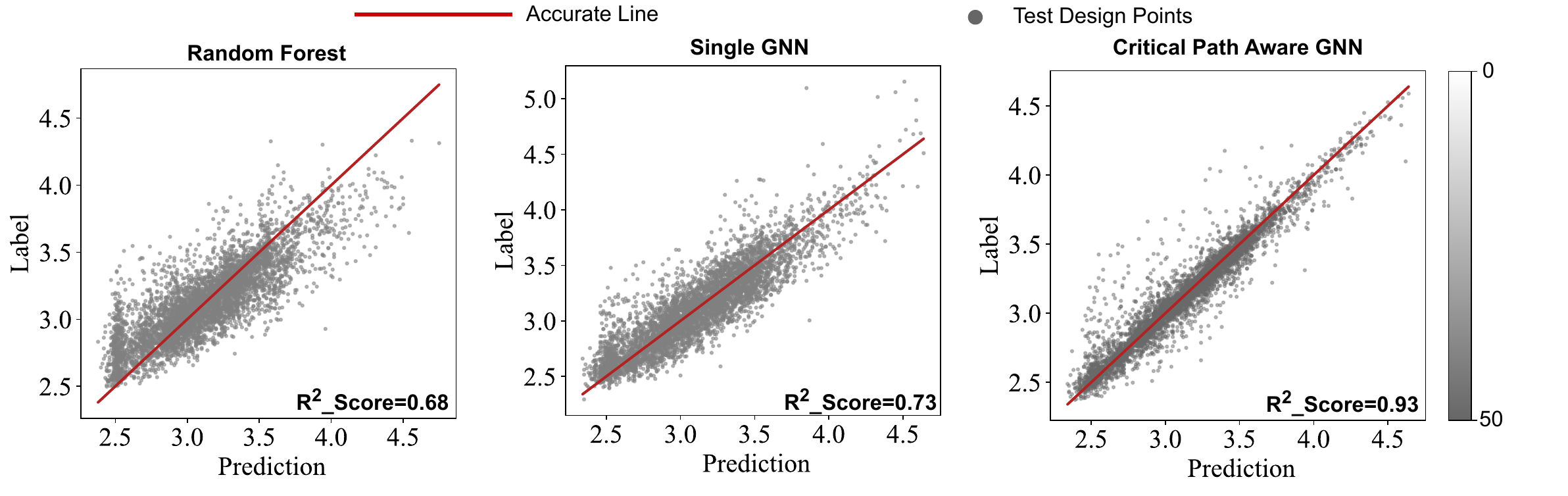}}
  \caption{Latency prediction model comparison among random forest, baseline single-stage GNN, and critical path aware two-stage GNN.}
  \label{fig:model-comparison}
\end{figure*}
\textbf{Graph Structure:} As mentioned, graph is the backbone of GNN model and we also investigate the influence of graph structure on the quality of the model,especially evaluate the efficiency of the proposed graph-based simplification strategy. Basically, we have both a naive graph that takes all the operators and non-arithmetic blocks as nodes and a simplified graph that merges the identical non-arithmetic nodes evaluated. We take K-means accelerator as an example and train GNN models based on both naive graph and the simplified graph. The prediction results are presented in Table \ref{tab:merge_model}. It can be seen that the  proposed node fusion strategy can efficiently merge identical/redundant nodes in the original graph without sacrificing model accuracy, further reducing the size of the graph.

\begin{table}[h]
\footnotesize
\centering
\Large
\caption{Naive graph vs. Simplified graph that has identical nodes merged.}
\label{tab:merge_model}
\resizebox{\linewidth}{!}{
\begin{tabular}{cc|ccccc}
\hline
\multicolumn{2}{c|}{Evaluation Metric}                       & Area   & Power  & Latency & SSIM  & Accuracy \\ \hline
\multicolumn{1}{c|}{\multirow{2}{*}{MAPE}}      & Naive      & 0.0120 & 0.0105 & 0.024   & 0.076 & 0.847    \\
\multicolumn{1}{c|}{}                           & Simplified & 0.0034 & 0.0099 & 0.019   & 0.053 & 0.88     \\ \hline
\multicolumn{1}{c|}{\multirow{2}{*}{R2\_Score}} & Naive      & 0.98   & 0.98   & 0.914   & 0.847 & 0.86     \\
\multicolumn{1}{c|}{}                           & Simplified & 0.99   & 0.98   & 0.925   & 0.86  & 0.88     \\ \hline
\end{tabular}
}
\end{table}

\textbf{GNN model architecture:} During the model construction phase, we need to select a Graph Neural Network (GNN) model for predicting and analyzing application circuits from different GNN models. However, there are several different GNN models architecture available. In this section, we utilize a dataset we've constructed to compare the predictive accuracy and performance of different GNN models architecture using the same training and testing sets.
 The table \ref{tab:different-gnn-model} shows the results of the Gauusion operator on different GNN models in terms of R2 coefficient. It is evident from the table that the GSAE model outperforms the other models significantly. Therefore, in the Critical Path Aware GNN Architecture, we adopt the GSAE model as the base model
\vspace{-0.5em}
\begin{table}[h]
\centering
 \caption{Comparison between different GNN models }
\resizebox{\linewidth}{!}{
\begin{tabular}{c|ccccc}
\hline
Model               & Area  & Power & Latency & SSIM  & Accuracy                             \\ \hline
GCN                 & 0.967 & 0.925 & 0.879   & 0.92  & 0.84                                 \\
MPNN                & 0.941 & 0.906 & 0.856   & 0.916 & 0.87                                 \\
GAT                 & 0.963 & 0.923 & 0.895   & 0.934 & 0.85                                 \\
GSAE                & 0.975 & 0.948 & 0.925    & 0.95  & 0.88                                  \\  \hline
\end{tabular}
}
\label{tab:different-gnn-model}
\end{table}
\subsection{Design Space Exploration}
\textbf{Selection of sampling methods:} There have been many DSE sampling algorithms that follow distinct search strategies over the design space. In this work, we take Sobel accelerator as an example and evaluate the different sampling algorithms, including random sampling, Bayesian\cite{malu2021bayesian}, NSGAII\cite{deb2002fast}, and NSGAIII\cite{mkaouer2015many}. Figure \ref{fig:different_sample_model} displays the Pareto curves constructed by different sampling methods for the Sobel operator with a sampling points of 1000. It can be observed that the NSGAIII based on genetic algorithms is significantly superior to other sampling methods. Therefore, in the design space exploration phase, we use NSGAIII as the design space sampling function.
\begin{figure}[t]
\centerline{\includegraphics[width=0.45\textwidth]{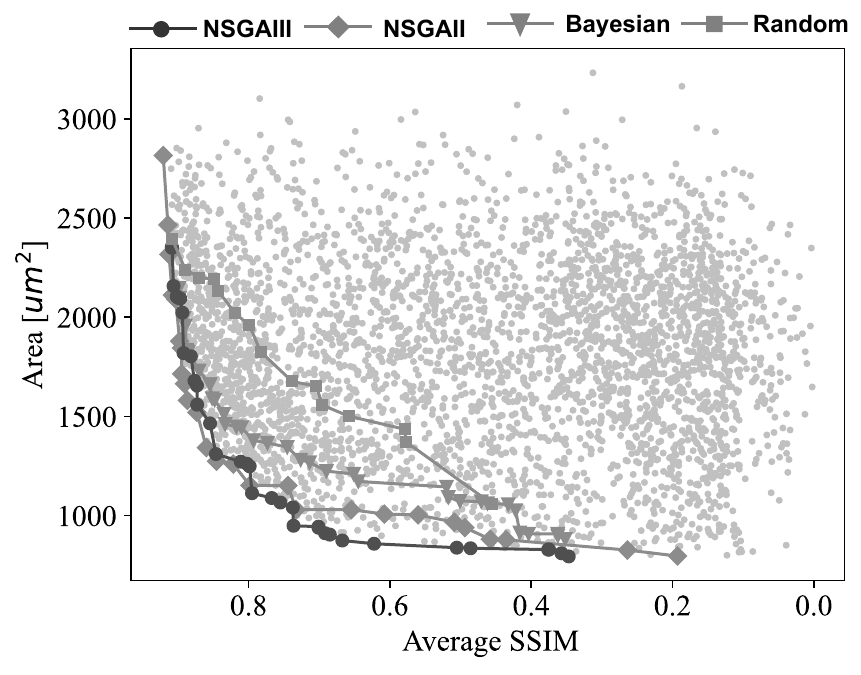}}
\caption{Comparison of Pareto curve construction using different sampling methods. }
\label{fig:different_sample_model}
\end{figure}

\textbf{Design Space Pruning:} An efficient design space pruning algorithm serves as the cornerstone for design space exploration. It not only facilitates a more uniform sampling of the training model\'s dataset but also enables the sampling of more beneficial points when constructing the Pareto curve.
We compared the original design space with the design space size after applying pruning algorithms. With an efficient pruning algorithm, we were able to eliminate redundant samples in the design space. Additionally, we validated the effectiveness of our pruning algorithm. When the design space expands, Autoax's pruning method tends to cause the clustering of sampled points around a certain value, thereby failing to guarantee the diversity of sampled points.As indicated by Table \ref{tab:design-space}, ApproxPilot demonstrates a more uniform construction of models compared to the Autoax\cite{mrazek2019autoax} method in latency and ssim.

\begin{table}[h]
\footnotesize
\centering
\Large
\caption{Design space of the approximate accelerators before and after pruning.}
\label{tab:design-space}
\resizebox{\linewidth}{!}{
\begin{tabular}{c|ccc}
\hline
\multirow{2}{*}{Operator} & \multirow{2}{*}{\begin{tabular}[c]{@{}c@{}}Initial\\ Space\end{tabular}} & \multirow{2}{*}{\begin{tabular}[c]{@{}c@{}}Invalid \\ Design Pruning\end{tabular}} & \multirow{2}{*}{\begin{tabular}[c]{@{}c@{}}Redundant \\ Design Pruning\end{tabular}} \\
                          &                                                                          &                                                                                    &                                                                                      \\ \hline
Sobel                     & $2.04\times10^{15}$                                                      & $8.56\times 10^{9}$                                                                & $2.56\times 10^{8}$                                                                  \\
Gaussian                  & $8.56\times10^{26}$                                                      & $5.64\times 10^{18}$                                                               & $9.32\times 10^{16}$                                                                 \\
Kmeans                    & $7.56\times10^{34}$                                                      & $6.42\times 10^{26}$                                                               & $5.82\times 10^{20}$                                                                 \\ \hline
\end{tabular}
}
\end{table}

\section{Conclusion}
In this work, we propose an end-to-end approximation optimization framework, ApproxPilot, for customized hardware accelerators. It can automatically select the suitable approximate arithmetic units from a vast design space for all the arithmetic units in the target accelerator while fulfilling multiple optimization objectives in terms of PPA and accuracy. The framework includes a series of dependent procedures including design space pruning, prediction model construction, and design space exploration, that are important to the overall approximation optimization. However, the key challenge of the design framework is the PPA and accuracy prediction that must be both fast and accurate to ensure an intensive design space exploration. Notable, we observe that critical path plays a key role in latency prediction and develop a GNN-based critical path aware prediction model, which significantly improves the prediction accuracy compared to straightforward GNN models and paves the way for automatic approximate accelerator optimization. According to our experiments on three different approximate accelerator optimization cases, ApproxPilot produces significantly better Pareto curves, demonstrating superior design trade-offs among PPA and accuracy than prior design frameworks.
\newpage
\bibliographystyle{IEEEtran}
\bibliography{myreferences}

\begin{thebibliography}{10}
\providecommand{\url}[1]{#1}
\csname url@samestyle\endcsname
\providecommand{\newblock}{\relax}
\providecommand{\bibinfo}[2]{#2}
\providecommand{\BIBentrySTDinterwordspacing}{\spaceskip=0pt\relax}
\providecommand{\BIBentryALTinterwordstretchfactor}{4}
\providecommand{\BIBentryALTinterwordspacing}{\spaceskip=\fontdimen2\font plus
\BIBentryALTinterwordstretchfactor\fontdimen3\font minus \fontdimen4\font\relax}
\providecommand{\BIBforeignlanguage}[2]{{%
\expandafter\ifx\csname l@#1\endcsname\relax
\typeout{** WARNING: IEEEtran.bst: No hyphenation pattern has been}%
\typeout{** loaded for the language `#1'. Using the pattern for}%
\typeout{** the default language instead.}%
\else
\language=\csname l@#1\endcsname
\fi
#2}}
\providecommand{\BIBdecl}{\relax}
\BIBdecl

\bibitem{jiang2020approximate}
H.~Jiang, F.~J.~H. Santiago, H.~Mo, L.~Liu, and J.~Han, ``Approximate arithmetic circuits: A survey, characterization, and recent applications,'' \emph{Proceedings of the IEEE}, vol. 108, no.~12, pp. 2108--2135, 2020.

\bibitem{xu2015approximate}
Q.~Xu, T.~Mytkowicz, and N.~S. Kim, ``Approximate computing: A survey,'' \emph{IEEE Design \& Test}, vol.~33, no.~1, pp. 8--22, 2015.

\bibitem{mittal2016survey}
S.~Mittal, ``A survey of techniques for approximate computing,'' \emph{ACM Computing Surveys (CSUR)}, vol.~48, no.~4, pp. 1--33, 2016.

\bibitem{mahdiani2009bio}
H.~R. Mahdiani, A.~Ahmadi, S.~M. Fakhraie, and C.~Lucas, ``Bio-inspired imprecise computational blocks for efficient vlsi implementation of soft-computing applications,'' \emph{IEEE Transactions on Circuits and Systems I: Regular Papers}, vol.~57, no.~4, pp. 850--862, 2009.

\bibitem{shin2011new}
D.~Shin and S.~K. Gupta, ``A new circuit simplification method for error tolerant applications,'' in \emph{2011 Design, Automation \& Test in Europe}.\hskip 1em plus 0.5em minus 0.4em\relax IEEE, 2011, pp. 1--6.

\bibitem{nepal2014abacus}
K.~Nepal, Y.~Li, R.~I. Bahar, and S.~Reda, ``Abacus: A technique for automated behavioral synthesis of approximate computing circuits,'' in \emph{2014 Design, Automation \& Test in Europe Conference \& Exhibition (DATE)}.\hskip 1em plus 0.5em minus 0.4em\relax IEEE, 2014, pp. 1--6.

\bibitem{mrazek2019autoax}
V.~Mrazek, M.~A. Hanif, Z.~Vasicek, L.~Sekanina, and M.~Shafique, ``autoax: An automatic design space exploration and circuit building methodology utilizing libraries of approximate components,'' in \emph{Proceedings of the 56th Annual Design Automation Conference 2019}, 2019, pp. 1--6.

\bibitem{prabakaran2023xel}
B.~S. Prabakaran, V.~Mrazek, Z.~Vasicek, L.~Sekanina, and M.~Shafique, ``Xel-fpgas: An end-to-end automated exploration framework for approximate accelerators in fpga-based systems,'' in \emph{2023 IEEE/ACM International Conference on Computer Aided Design (ICCAD)}.\hskip 1em plus 0.5em minus 0.4em\relax IEEE, 2023, pp. 1--9.

\bibitem{dai2018fast}
S.~Dai, Y.~Zhou, H.~Zhang, E.~Ustun, E.~F. Young, and Z.~Zhang, ``Fast and accurate estimation of quality of results in high-level synthesis with machine learning,'' in \emph{2018 IEEE 26th Annual International Symposium on Field-Programmable Custom Computing Machines (FCCM)}.\hskip 1em plus 0.5em minus 0.4em\relax IEEE, 2018, pp. 129--132.

\bibitem{wu2021ironman}
N.~Wu, Y.~Xie, and C.~Hao, ``Ironman: Gnn-assisted design space exploration in high-level synthesis via reinforcement learning,'' in \emph{Proceedings of the 2021 on Great Lakes Symposium on VLSI}, 2021, pp. 39--44.

\bibitem{wu2022high}
N.~Wu, H.~Yang, Y.~Xie, P.~Li, and C.~Hao, ``High-level synthesis performance prediction using gnns: Benchmarking, modeling, and advancing,'' in \emph{Proceedings of the 59th ACM/IEEE Design Automation Conference}, 2022, pp. 49--54.

\bibitem{li2022noception}
F.~Li, Y.~Wang, C.~Liu, H.~Li, and X.~Li, ``Noception: A fast ppa prediction framework for network-on-chips using graph neural network,'' in \emph{2022 Design, Automation \& Test in Europe Conference \& Exhibition (DATE)}.\hskip 1em plus 0.5em minus 0.4em\relax IEEE, 2022, pp. 1035--1040.

\bibitem{wu2022ironman}
N.~Wu, Y.~Xie, and C.~Hao, ``Ironman-pro: Multiobjective design space exploration in hls via reinforcement learning and graph neural network-based modeling,'' \emph{IEEE Transactions on Computer-Aided Design of Integrated Circuits and Systems}, vol.~42, no.~3, pp. 900--913, 2022.

\bibitem{10416120}
H.~Kuang, X.~Cao, J.~Li, and L.~Wang, ``Hgbo-dse: Hierarchical gnn and bayesian optimization based hls design space exploration,'' in \emph{2023 International Conference on Field Programmable Technology (ICFPT)}, 2023, pp. 106--114.

\bibitem{scarabottolo2020approximate}
I.~Scarabottolo, G.~Ansaloni, G.~A. Constantinides, L.~Pozzi, and S.~Reda, ``Approximate logic synthesis: A survey,'' \emph{Proceedings of the IEEE}, vol. 108, no.~12, pp. 2195--2213, 2020.

\bibitem{scarabottolo2018circuit}
I.~Scarabottolo, G.~Ansaloni, and L.~Pozzi, ``Circuit carving: A methodology for the design of approximate hardware,'' in \emph{2018 Design, Automation \& Test in Europe Conference \& Exhibition (DATE)}.\hskip 1em plus 0.5em minus 0.4em\relax IEEE, 2018, pp. 545--550.

\bibitem{brent1982regular}
Brent and Kung, ``A regular layout for parallel adders,'' \emph{IEEE transactions on Computers}, vol. 100, no.~3, pp. 260--264, 1982.

\bibitem{han1987fast}
T.~Han and D.~A. Carlson, ``Fast area-efficient vlsi adders,'' in \emph{1987 IEEE 8th symposium on computer arithmetic (ARITH)}.\hskip 1em plus 0.5em minus 0.4em\relax IEEE, 1987, pp. 49--56.

\bibitem{kahng2012accuracy}
A.~B. Kahng and S.~Kang, ``Accuracy-configurable adder for approximate arithmetic designs,'' in \emph{Proceedings of the 49th annual design automation conference}, 2012, pp. 820--825.

\bibitem{liu2015design}
W.~Liu, L.~Chen, C.~Wang, M.~O’Neill, and F.~Lombardi, ``Design and analysis of inexact floating-point adders,'' \emph{IEEE Transactions on Computers}, vol.~65, no.~1, pp. 308--314, 2015.

\bibitem{rehman2016architectural}
S.~Rehman, W.~El-Harouni, M.~Shafique, A.~Kumar, J.~Henkel, and J.~Henkel, ``Architectural-space exploration of approximate multipliers,'' in \emph{2016 IEEE/ACM International Conference on Computer-Aided Design (ICCAD)}.\hskip 1em plus 0.5em minus 0.4em\relax IEEE, 2016, pp. 1--8.

\bibitem{saadat2019approximate}
H.~Saadat, H.~Javaid, and S.~Parameswaran, ``Approximate integer and floating-point dividers with near-zero error bias,'' in \emph{Proceedings of the 56th Annual Design Automation Conference 2019}, 2019, pp. 1--6.

\bibitem{shafique2015low}
M.~Shafique, W.~Ahmad, R.~Hafiz, and J.~Henkel, ``A low latency generic accuracy configurable adder,'' in \emph{Proceedings of the 52nd Annual Design Automation Conference}, 2015, pp. 1--6.

\bibitem{kulkarni2011trading}
P.~Kulkarni, P.~Gupta, and M.~Ercegovac, ``Trading accuracy for power with an underdesigned multiplier architecture,'' in \emph{2011 24th Internatioal Conference on VLSI Design}.\hskip 1em plus 0.5em minus 0.4em\relax IEEE, 2011, pp. 346--351.

\bibitem{mrazek2017evoapprox8b}
V.~Mrazek, R.~Hrbacek, Z.~Vasicek, and L.~Sekanina, ``Evoapprox8b: Library of approximate adders and multipliers for circuit design and benchmarking of approximation methods,'' in \emph{Design, Automation \& Test in Europe Conference \& Exhibition (DATE), 2017}.\hskip 1em plus 0.5em minus 0.4em\relax IEEE, 2017, pp. 258--261.

\bibitem{ullah2018smapproxlib}
S.~Ullah, S.~S. Murthy, and A.~Kumar, ``Smapproxlib: Library of fpga-based approximate multipliers,'' in \emph{Proceedings of the 55th Annual Design Automation Conference}, 2018, pp. 1--6.

\bibitem{10131995}
K.~Zhong, S.~Zeng, W.~Hou, G.~Dai, Z.~Zhu, X.~Zhang, S.~Xiao, H.~Yang, and Y.~Wang, ``Cognn: An algorithm-hardware co-design approach to accelerate gnn inference with minibatch sampling,'' \emph{IEEE Transactions on Computer-Aided Design of Integrated Circuits and Systems}, vol.~42, no.~12, pp. 4883--4896, 2023.

\bibitem{10411398}
Z.~Qu, D.~Niu, S.~Li, H.~Zheng, and Y.~Xie, ``Tt-gnn: Efficient on-chip graph neural network training via embedding reformation and hardware optimization,'' in \emph{2023 56th IEEE/ACM International Symposium on Microarchitecture (MICRO)}, 2023, pp. 452--464.

\bibitem{9401612}
R.~Guirado, A.~Jain, S.~Abadal, and E.~Alarcón, ``Characterizing the communication requirements of gnn accelerators: A model-based approach,'' in \emph{2021 IEEE International Symposium on Circuits and Systems (ISCAS)}, 2021, pp. 1--5.

\bibitem{10476462}
F.~Niu, J.~Yue, J.~Shen, X.~Liao, and H.~Jin, ``Flashgnn: An in-ssd accelerator for gnn training,'' in \emph{2024 IEEE International Symposium on High-Performance Computer Architecture (HPCA)}, 2024, pp. 361--378.

\bibitem{ustun2020accurate}
E.~Ustun, C.~Deng, D.~Pal, Z.~Li, and Z.~Zhang, ``Accurate operation delay prediction for fpga hls using graph neural networks,'' in \emph{Proceedings of the 39th international conference on computer-aided design}, 2020, pp. 1--9.

\bibitem{yazdanbakhsh2016axbench}
A.~Yazdanbakhsh, D.~Mahajan, H.~Esmaeilzadeh, and P.~Lotfi-Kamran, ``Axbench: A multiplatform benchmark suite for approximate computing,'' \emph{IEEE Design \& Test}, vol.~34, no.~2, pp. 60--68, 2016.

\bibitem{deb2013evolutionary}
K.~Deb and H.~Jain, ``An evolutionary many-objective optimization algorithm using reference-point-based nondominated sorting approach, part i: solving problems with box constraints,'' \emph{IEEE transactions on evolutionary computation}, vol.~18, no.~4, pp. 577--601, 2013.

\bibitem{arbelaez2010contour}
P.~Arbelaez, M.~Maire, C.~Fowlkes, and J.~Malik, ``Contour detection and hierarchical image segmentation,'' \emph{IEEE transactions on pattern analysis and machine intelligence}, vol.~33, no.~5, pp. 898--916, 2010.

\bibitem{vasicek2014evolutionary}
Z.~Vasicek and L.~Sekanina, ``Evolutionary approach to approximate digital circuits design,'' \emph{IEEE Transactions on Evolutionary Computation}, vol.~19, no.~3, pp. 432--444, 2014.

\bibitem{ash1999r2}
A.~Ash and M.~Shwartz, ``R2: a useful measure of model performance when predicting a dichotomous outcome,'' \emph{Statistics in medicine}, vol.~18, no.~4, pp. 375--384, 1999.

\bibitem{tayman1999validity}
J.~Tayman and D.~A. Swanson, ``On the validity of mape as a measure of population forecast accuracy,'' \emph{Population Research and Policy Review}, vol.~18, pp. 299--322, 1999.

\bibitem{malu2021bayesian}
M.~Malu, G.~Dasarathy, and A.~Spanias, ``Bayesian optimization in high-dimensional spaces: A brief survey,'' in \emph{2021 12th International Conference on Information, Intelligence, Systems \& Applications (IISA)}.\hskip 1em plus 0.5em minus 0.4em\relax IEEE, 2021, pp. 1--8.

\bibitem{deb2002fast}
K.~Deb, A.~Pratap, S.~Agarwal, and T.~Meyarivan, ``A fast and elitist multiobjective genetic algorithm: Nsga-ii,'' \emph{IEEE transactions on evolutionary computation}, vol.~6, no.~2, pp. 182--197, 2002.

\bibitem{mkaouer2015many}
W.~Mkaouer, M.~Kessentini, A.~Shaout, P.~Koligheu, S.~Bechikh, K.~Deb, and A.~Ouni, ``Many-objective software remodularization using nsga-iii,'' \emph{ACM Transactions on Software Engineering and Methodology (TOSEM)}, vol.~24, no.~3, pp. 1--45, 2015.

\end{thebibliography}

%

\end{document}